\documentclass[aps,prb,twocolumn,superscriptaddress]{revtex4}
\usepackage{graphicx}
\usepackage{amsmath,amssymb}
\usepackage{times}
\def\v#1{\mathbf{#1}}
\def\nn{\nonumber}

\begin{document}


\title{Quantum Nernst Effect in a Bismuth Single Crystal}


\author{M. Matsuo}\email[]{matsuom@imr.tohoku.ac.jp}
\affiliation{Graduate School of Humanities and Sciences, Ochanomizu University, 
Otsuka, Bunkyo, Tokyo 112-8610,  Japan}
\affiliation{Institute of Industrial Science, University of Tokyo, Komaba, Meguro, Tokyo 153-8505, Japan}

\author{A. Endo}
\affiliation{Institute for Solid State Physics, University of Tokyo, Kashiwanoha, Kashiwa, Chiba 277-8581,  Japan}

\author{N. Hatano}
\affiliation{Institute of Industrial Science, University of Tokyo, Komaba, Meguro, Tokyo 153-8505, Japan}

\author{H. Nakamura}
\affiliation{Department of Simulation Science, National Institute for Fusion Science, Oroshi-cho, Toki, Gifu  509-5292, Japan}

\author{R. Shirasaki}
\affiliation{Department of Physics, Yokohama National University, 79-5 Tokiwadai, Yokohama 240-8501, Japan}

\author{K. Sugihara}
\affiliation{1-40-6-506 Shibayama, Funabashi, Chiba 274-0816, Japan}

\date{\today}

\begin{abstract}
We calculate the phonon-drag contribution to the transverse (Nernst) thermoelectric power $S_{yx}$ in a bismuth single crystal subjected to a quantizing magnetic field. The calculated heights of the Nernst peaks originating from the hole Landau levels and their temperature dependence reproduce the right order of magnitude for those of the pronounced magneto-oscillations recently reported by Behnia {\it et al}
[Phys.~Rev.~Lett.~{\bf 98}, 166602 (2007)]. 
A striking experimental finding that $S_{yx}$ is much larger than the longitudinal (Seebeck) thermoelectric power $S_{xx}$ can be naturally  explained as the effect of the phonon drag, combined with the well-known relation between the longitudinal and the Hall resistivity $\rho_{xx}\gg|\rho_{yx}|$ in a semi-metal bismuth.  
The calculation that includes the contribution of both holes and electrons suggests that some of the hitherto unexplained minor peaks located roughly at the fractional filling of the hole Landau levels are attributable to the electron Landau levels.
\end{abstract}


\maketitle

\section{Introduction}

A semi-metal bismuth has been attracting longstanding interest in the solid-state physics owing to its fascinating properties. The extraordinarily low carrier densities ($\sim\!\!\! 10^{-5}$ per atom) and small effective masses ($\sim\!\!\! 10^{-2} m_0$ with $m_0$ the free electron mass) combined with the availability of high-quality single crystals with highly mobile carriers render it an archetypal material for investigating the phenomena originating from the Landau quantization.
In fact, a plethora of magneto-oscillation phenomena, including the de Hass-van Alphen and the Shubnikov-de Hass effects, were first discovered in bismuth,~\cite{EdelmanReview}
 illustrating distinguished roles played by the material in the history of the solid-state physics.
Bismuth remains to be a subject of intensive ongoing studies spurred by its intriguing properties 
such as multi-valley degeneracy of Dirac-type electrons,~\cite{LuLi} enhanced spin-orbit interaction on the surface,~\cite{Hirahara} strong diamagnetism advantageous for the potential observation of the quantum spin-Hall effect.~\cite{Murakami,Fu}

The target of the present paper is the thermoelectric response of bismuth in a quantizing magnetic field.
In a magnetic field $\boldsymbol{B}$ applied perpendicular to the temperature gradient $\boldsymbol{\nabla} T$, the thermopower tensor contains not only the longitudinal (Seebeck effect, $S_{xx}$) but also the transverse (Nernst effect, $S_{yx}$) components, where we set the direction of $\boldsymbol{\nabla} T$ and $\boldsymbol{B}$ as the $x$ and $z$ directions, respectively. It is worth mentioning that the Nernst effect was also originally discovered in bismuth.~\cite{NernstEttings}
Magneto-oscillations of $S_{xx}$ and $S_{yx}$ due to the Landau quantization have been extensively studied in two-dimensional electron gases (2DEGs).~\cite{Fletcher} The effect of the Landau quantization is expected to be less easily observed in three-dimensional (3D) materials. 
Nevertheless, the initial observation of the magneto-oscillation in the thermoelectric coefficients of bismuth dates back to several decades ago.~\cite{Steele,Mangez,Farag} The thermopower of bismuth has attracted renewed interest since the publication of recent experimental works by Behnia {\it et al.}~\cite{Behnia,BehniaScience}
They extended the measurement to lower temperatures ($\sim\!\! 0.3$ K) and higher magnetic fields ($\sim\!\! 30$ T) and reported prominent magneto-oscillations that rather appear as a series of discrete peaks ~\cite{Behnia}
 and further, small features in the ultraquantum limit that possibly signals the fractional quantization in three dimensions.~\cite{BehniaScience} The oscillations in the thermopower were much more pronounced than the oscillations in the resistivity (the Shubnikov-de Haas oscillations).
Interestingly, the Nernst signal $S_{yx}$ was found to be much larger than the Seebeck signal $S_{xx}$ in bismuth, in marked contrast to the case in 2DEGs, where generally $S_{xx}>|S_{yx}|$. Moreover, the line shape of $S_{yx}$ in bismuth was quite unlike that in 2DEGs: the former takes a peak when the chemical potential crosses a Landau level (as is the case in $S_{xx}$ for 2DEGs), while in 2DEGs, $S_{yx}$ changes sign.~\cite{Fletcher} The amplitudes of the peaks were large ($\sim$mV/K), and the peak heights rapidly increased with temperature. These findings, as well as the origin of small peaks located between the main peaks attributable to the Landau levels of the holes, remain unexplained. In an initial attempt toward the understanding, the present authors extended to 3D the theory for 2DEGs by Nakamura \textit{et al.}~\cite{nernst} that invokes the edge-current picture.~\cite{ISQM}
Although the calculation qualitatively reproduced the main peaks of the experimental traces, the amplitudes were found to be orders of magnitude smaller ($\sim$ 10 $\mu$V/K). 
Furthermore, the theory failed to reproduce the strong temperature dependence. 
Note that the thermopower originating from the edge current corresponds to the contribution of the carrier diffusion in the clean limit in a quantizing magnetic field.~\cite{nernst,Jonson84,Oji84} Inclusion of disorders was shown to further reduce the magnitude.~\cite{Jonson84,shirasaki} Therefore, the experimentally observed large-amplitude oscillation is not attributable to the diffusion contribution.

In the present paper, we show that the large amplitude, the temperature dependence, and the dominance of $S_{yx}$ over $S_{xx}$ can be consistently explained as the effect of the phonon drag in the system containing both holes and electrons as carriers.
Note that the phonon-drag contribution is known to play a dominant role also in 2DEGs.~\cite{Fletcher}
Preliminary results of the phonon-drag contribution that consider only holes as carriers were already presented in Ref.~\onlinecite{ISQM}.  
Here we describe more refined calculation that takes account of contributions of electrons, the charge neutral condition, and the Zeeman splitting neglected in Ref.~\onlinecite{ISQM}. 
The calculation suggests that the minor peaks that appear at locations where fractional numbers of the hole Landau levels are filled actually originate from electron Landau levels.


\section{Phonon-drag contribution to transverse thermopower}

The Hamiltonian of the system with a magnetic field $\boldsymbol{B}$ and a small electric field $F_y$ applied in the $z$ (trigonal axis of bismuth) and $y$ directions, repectively, is given by
\begin{equation}
\label{eq-ham}
\mathcal{H}_a
={ \frac{1}{2}}(\boldsymbol{p}-e_a\boldsymbol{A}){\mathbb{M}_a}^{-1}
(\boldsymbol{p}-e_a\boldsymbol{A})
+\sigma g_a\mu_B B - e_a F_y y,
\end{equation}
where $\boldsymbol{A}=(-B y, 0, 0)$ denotes the vector potential and $\sigma=\pm 1/2$ the spin. The suffix $a$ is used throughout the paper to indicate the quantity either of a hole ($a=\mathrm{h}$) or of an electron ($a=\mathrm{e}$), with $e_\mathrm{h}=e$ and $e_\mathrm{e}=-e$ ($e>0$).
The effective mass tensors for holes and electrons are 
\begin{eqnarray}
\mathbb{M}_{\rm h} = 
\left(
\begin{array}{ccc}
m_{\mathrm{h}x} & 0 & 0 \\
0 & m_{\mathrm{h}x} & 0 \\
0 & 0 & m_{\mathrm{h}z}
\end{array}
\right) 
\end{eqnarray}
and 
\begin{eqnarray}
\mathbb{M}_{\rm e} = 
\left(
\begin{array}{ccc}
m_{\mathrm{e}x} & 0 & m_{\mathrm{e}xz} \\
0 & m_{\mathrm{e}y} & 0 \\
m_{\mathrm{e}xz} & 0 & m_{\mathrm{e}z}
\end{array}
\right),
\end{eqnarray}
respectively.~\cite{elemass} The values of the components are listed in Table~\ref{param}.
\begin{table}[b]
\caption{Parameter values used in our calculation, taken from Refs.~\onlinecite{effectiveMass,Zeeman,deform}.
We calculated the Zeeman energy of electrons by Smith's method (Ref.~\onlinecite{smith}) using the effective masses in Ref.~\onlinecite{effectiveMass}.}
\begin{tabular}{lcc}
\hline
 & Hole & Electron  \\ \hline
Effective mass ($m_0$) ~\cite{effectiveMass}  &$m_{{\rm h}x}=0.06289$&$m_{{\rm e}x}=0.26$  \\ \cline{2-3}
 & --- &$m_{{\rm e}y}=0.00113   \,$ \\ \cline{2-3}
&$m_{{\rm h}z}=0.6667$ &$m_{{\rm e}z}=0.00443$ \\  \cline{2-3}
& --- &$m_{{\rm e}xz}=-0.0195$ \\ \hline
Zeeman energy $g \mu_B$&$2.16\,\hbar \omega_{\rm h}$ ~\cite{Zeeman}&$0.5849 \, \hbar \omega_{\rm e}$ ~\cite{smith}\\  \hline
Deformation potential ~\cite{deform}&  $D_{\rm h}=1.2$ eV &  $D_{\rm e}=2.2$ eV \\  \hline
\multicolumn{3}{c}{Band gap at L point $E_g=15.3$ meV ~\cite{effectiveMass}} \\
\multicolumn{3}{c}{Band overlap $\mu_0=38.5$ meV ~\cite{effectiveMass}}  \\
\multicolumn{3}{c}{Group velocity of phonons $v_s=2 \times 10^3 \, {\rm m/s}$ ~\cite{deform}} \\
\multicolumn{3}{c}{Density $\rho=9.75\times 10^3  \, {\rm kg/m^3}$}\\ 
\multicolumn{3}{c}{Size of the sample $W=2.2$ mm, $L=4.0$ mm ~\cite{Behnia}}\\ \hline
\end{tabular} 
\label{param}
\end{table}
The eigenenergy of the Hamiltonian (\ref{eq-ham}) in first order of $F_y$ reads
\begin{equation}
\label{eq-energy}
E_a(n,k_z,\sigma)=\hbar\omega_a\left(n+\frac{1}{2}\right)
+\frac{\hbar^2{k_z}^2}{2m_{az}}+\sigma g_a\mu_B B - e_a F_y Y_{0a}
\end{equation}
with the cyclotron frequency $\omega_a\equiv e B/m_{a}$, 
where the cyclotron mass $m_{a}$ is given by $m_{\mathrm{h}}=m_{\mathrm{h}x}$ and $m_{\mathrm{e}}=\sqrt{\det\mathbb{M}_\mathrm{e}/m_{\mathrm{e}z}}$.~\cite{smith,Lax2}
The corresponding eigenfunction is
$\psi_a(y-Y_{0a};n,k_x,k_z)=\phi_a(y-Y_{0a};n) \exp[i(k_xx+k_zz)]$,
where
$Y_{0 {\rm h}}= \hbar k_x/(e B)$,
$Y_{0 {\rm e}}= -\hbar (k_x - k_z m_{{\rm e}xz}/m_{{\rm e}z})/(e B)$,
and
\begin{equation}
\phi_a(y;n)\equiv\left(2^nn!\sqrt{\pi}l_a\right)^{-1/2}
e^{-y^2/(2{l_a}^2)}H_n(y/l_a),
\end{equation}
with the magnetic length $l_a = \sqrt{\hbar / (m_{ay} \omega_a})$ represented as $l_\mathrm{h}=\sqrt{\hbar/(eB)}$ and $l_\mathrm{e}=\sqrt{m_\mathrm{e}/m_{\mathrm{e}y}}\sqrt{\hbar/(eB)}$. 
 
We now describe our calculation of the phonon-drag effect.
The phonon-drag thermopower in a magnetic field was studied for bismuth by Sugihara ~\cite{Sugihara,Sugihara2} and for a  GaAs/AlGaAs 2DEG by Kubakaddi {\it et al}.~\cite{Kubakaddi}
We here closely follow Sugihara's calculation.
The difference from his calculation is that we treat the Fermi and Bose distributions exactly and evaluate the magnetic-field dependence numerically.
For  the calculation of the thermopower, 
there are two equivalent approaches.
In the $Q$ approach, we calculate the electric current under a temperature gradient, 
while in the $\Pi$ approach, we calculate the heat current under an electric field.
The two approaches are related through the Kelvin-Onsager relation;~\cite{Herring} 
$S_{yx}(\boldsymbol{B}) = \Pi_{xy}(- \boldsymbol{B})/T  \label{Kelvin}$, 
where $\Pi_{xy}$ is the Peltier coefficient.
Here we follow the $\Pi$ approach.
Carriers accelerated by the electric field $F_y$ ``drag''  phonons because of carrier-phonon interaction and thus generate the heat current of phonons. The heat currents of holes and electrons are negligibly smaller than that of phonons.
Then the Peltier coefficient is given by $\Pi_{xy}=  Q_x \rho_{xx}/F_y$, 
where $Q_x$ denotes the heat current of phonons in the $x$ direction and we used the relation $\rho_{xx} \gg |\rho_{yx}|$ characteristic of the systems that contain both holes and electrons as carriers, where $\rho_{xx}$ and $\rho_{yx}$ denote the longitudinal and the Hall resistivities, respectively.
\noindent
At low temperatures  we may neglect all lattice excitations except 
 acoustic phonons with the energy $\hbar \omega_q $ and the wave vector $\mbox{\boldmath $q$}$, which are generated through deformation coupling.
The heat current of phonons in the $x$ direction is then given by
\begin{equation}
Q_x = \int \frac{d \mbox{\boldmath $q$}}{(2\pi)^3}  \hbar \omega_q v_s \frac{q_x}{q} 
g(\mbox{\boldmath $q$}) , 
\label{Qy}
\end{equation}
where $\omega_q=v_s q$, $v_s$ is the group velocity of the phonons
 and 
 $g(\mbox{\boldmath $q$})= N_{\mbox{\boldmath $q$}} - N_q^{(0)} $ 
 represents the displacement of the phonon distribution $N_{\mbox{\boldmath $q$}}$ from its equilibrium Bose distribution $N_q^{(0)}$.
In order to estimate the displacement, 
we use the Boltzmann equation in the steady state;
\begin{equation}
\left( \frac{\partial N_{\mbox{\boldmath $q$}}}{\partial t} \right)_{\rm carrier} + 
\left( \frac{\partial N_{\mbox{\boldmath $q$}}}{\partial t} \right)_{\rm relaxation} = 0 .
\label{Boltzmann}
\end{equation}
The first term of the left-hand side represents the change in the phonon distribution due to interaction with carriers and the second term represents that due to other interactions such as boundary scattering, phonon-phonon interaction and impurity scattering.
These two terms are balanced in the steady state.

We estimate the quantity $(\partial N_{\mbox{\boldmath $q$}}/\partial t)_{{\rm carrier}}$ 
in the Born approximation as
\begin{eqnarray}
\biggl(\frac{\partial N_{\mbox{\boldmath $q$}}}{\partial t}\biggr)_{{\rm carrier}}
&=& \sum_{\alpha, \alpha'}
 \biggl[
W^{({\rm em})} {(\alpha', \alpha)} \, f_{\alpha'} (1 - f_{\alpha}) 
\nonumber \\
& &\hspace{0.5cm} 
- W^{({\rm ab})}{(\alpha, \alpha')} \, f_{\alpha} (1 - f_{\alpha'})
\biggr], \label{abem}
\end{eqnarray}
where 
$f_{\alpha}=f\left[ E(\alpha) \right]$ is the Fermi distribution of carriers in a state $\alpha$.
Each $\alpha$ represents the set of three quantum numbers
$(n,k_x,k_z)$, and
$W^{({\rm em})}(\alpha',\alpha)$ and $W^{({\rm ab})}(\alpha,\alpha')$ are the transition probabilities
 from a state $\alpha$ to a state $\alpha'$ by emitting or absorbing a phonon, respectively, given by 
Fermi's golden rule, 
\begin {eqnarray}
 \begin{array}{l}
W^{(\rm{em})}=  N_{\mbox{\boldmath $q$}}+1 \\
W^{(\rm{ab})}=  N_{\mbox{\boldmath $q$}}
 \end{array}
\Biggr\} \times
\frac{2\pi \left|V_q\right|^2}{\hbar} 
\left| \langle \psi_{\alpha'} |e^{\pm i \v{q}\cdot \v{r}}|\psi_{\alpha} \rangle
\right|^2 \nn \\
 \times \delta \left( E_a{(\alpha')} - E_a{(\alpha)} - \hbar \omega_q \right)
 \label{fermigolden}
\end{eqnarray}
with
$\left|V_q\right|^2 = D_a^2 {\hbar q}/{(2 \rho V v_s)}$, 
where $\rho,V$, and $D_a$ are the bismuth density, the sample volume, and the deformation potential of carriers, respectively. 
Expanding Eq.~(\ref{abem}) in O$\left({F_y}\right)$ we have the first term in Eq.~(\ref{Boltzmann}).
 In the second term, 
we use  the relaxation-time approximation;
$(\partial N_{\mbox{\boldmath $q$}} / \partial t)_{{\rm relaxation}}
 = - g(\mbox{\boldmath $q$}) /\tau_ r(\mbox{\boldmath $q$})$. 
The carrier-phonon interaction changes the phonon distribution, but 
other interactions make the nonequilibrium distribution relax back to the equilibrium one in time 
$\tau_r$.
Solving Eq.~(\ref{Boltzmann}) with respect to $g(\mbox{\boldmath $q$})$,
we obtain
\begin{eqnarray}
g(\mbox{\boldmath $q$}) = - \frac{\hbar F_y \tilde{q}_a}{k_B T B}N_q^{(0)}
 (N_q^{(0)}+1)
  \frac{\tau_{\rm{tot}}(\mbox{\boldmath $q$})}{\tau_{\rm c}(\mbox{\boldmath $q$})}, \label{Nqdiff}
\end{eqnarray}
where
\begin{eqnarray}
& &\frac{1}{\tau_{\rm c}(\mbox{\boldmath $q$})}
\equiv \frac{2\pi}{\hbar} |V_q|^2 \sum_{\alpha,\alpha'} 
|\langle \psi_a(\alpha') | e^{- i \v{q}\cdot \v{r}}| \psi_a(\alpha) \rangle|^2
\nonumber \\
& &\hspace{0.5cm}\times
\frac{f_{\alpha}(1 - f_{\alpha'})}{N_q^{(0)}+1}  
 \delta(E_a(\alpha') - E_a(\alpha) - \hbar \omega_q),
 \label{1ptauc}
  \end{eqnarray}
${\tau_{\rm{tot}}(\mbox{\boldmath $q$})}^{-1}
={\tau_r(\mbox{\boldmath $q$})}^{-1}+{\tau_{\rm {c}}(\mbox{\boldmath $q$})}^{-1}$, and 
  $\tilde{q}_{\rm h}=q_x$, $\tilde{q}_{\rm e}=q_x - q_z m_{{\rm e}xz}/m_{{\rm e}z}$. 
At low temperatures, the phonons in a bismuth single crystal are known to be ballistic and the boundary scattering is dominant,~\cite{Sugihara,Farag,Behnia07-1} and therefore
we set ${\tau_{\rm{tot}}(\mbox{\boldmath $q$})}^{-1} \cong v_s/L$, where $L$ is the length in the $x$ direction.
By plugging Eq.~(\ref{Nqdiff}) into Eq.~(\ref{Qy}), we have
\begin{eqnarray}
\frac{Q_{x a}}{F_y} = - \frac{1}{(2 \pi)^4} \frac{e \hbar L D_a^2}{2 k_B T \rho}
\sum_{n,n'} \int {\rm d} \mbox{\boldmath $q$} \,  q_x \tilde{q}_a \, q \, N^{(0)}_{q} I_{a n,n'}(\mbox{\boldmath $q$}) 
\nn \\
\times
\int{\rm d}k_z f_{\alpha} (1 - f_{\alpha'}) \delta \left( E_a({\alpha'}) - E_a({\alpha}) - \hbar \omega_q
 \right), \label{Qydz}
\end{eqnarray}
where
\begin{eqnarray}
&&I_{a n,n'}(\mbox{\boldmath $q$}) \nn \\
& &= \left|
\int_{-\infty}^{\infty}   \phi_{a}(y - \hbar \tilde{q}_a/(e_aB);n') e^{- i q_y y}
 \phi_{a}(y;n) {\rm d}y \right|^2.
\end{eqnarray}
We thus arrive at $S_{yx}= - Q_x \rho_{xx}/(F_y T)$ by adding Eq.\ (\ref{Qydz}) up over the spin degree of freedom and also over electrons and holes. The integration with respect to $k_z$ can be done analytically, and we obtain
\begin{eqnarray}
S_{yx}& &= - \frac{1}{(2 \pi)^4} 
\frac{e \rho_{xx} L}{2 k_B T^2 \rho \hbar} 
\sum_{a={\rm h,e}} D_{a}^2 m_{az}^*
\sum_{\sigma} \sum_{n,n'}   \nn \\
& &\times \,  \int {\rm d}\mbox{\boldmath $q$} \, \frac{q_x \tilde{q}_a \, q \,}{q_z} 
 N^{(0)}_{q} 
 I_{a n,n'} (\mbox{\boldmath $q$}) \nn \\
& &\times f\left[E_a(n,k_{z0 a},\sigma)\right]\left\{1 - f\left[E_a(n',k_{z0 a}+q_z,\sigma)\right]\right\} \nn \\
& & 
 \label{syxlast}
\end{eqnarray}
with
\begin{equation}
k_{z0 a}=\frac{m_{az}}{\hbar q_z} [\omega_a (n-n')+\omega_q]-\frac{1}{2}q_z.
\label{kz0}
\end{equation}
Finally the integration with respect to $\mbox{\boldmath $q$}$ is performed numerically. For the values of $\rho_{xx}$, we made use of the experimental data (at 0.25 K) by Behnia.~\cite{Behniadata}

\section{Results of calculation and comparison with experiment}

\begin{figure}
\begin{center}
\includegraphics[width=3.2in]{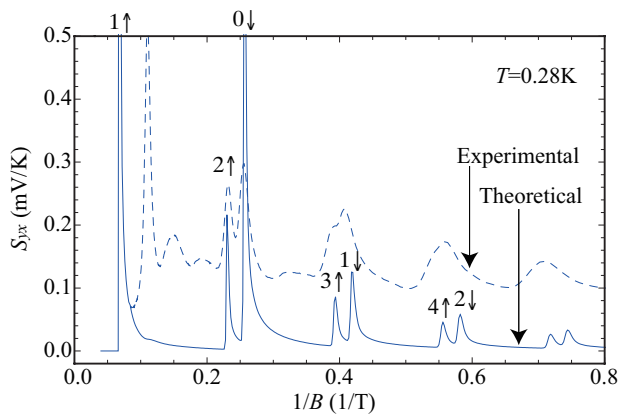}
\end{center}
\caption{(Color online).  The transverse thermopower $S_{yx}$ of holes at $T$= 0.28 K against the inverse magnetic field $1/B$.
 The solid lines are our theoretical results and the broken lines are the experimental results by Behnia {\it et al.}~\protect\cite{Behnia}
The peaks are labeled by $(n,{\rm spin})$.}
\label{phonondraghole}
\end{figure}

We first consider only holes as carriers; holes produce a greater contribution than electrons because the effective mass in the direction of the magnetic field, hence the density-of-states peak at a Landau level, is larger for holes than for electrons.
In Fig.~\ref{phonondraghole}  we compare the theoretical and experimental results at $T=0.28$ K.  In the calculation we used the parameter values given in Table~\ref{param} and the constant chemical potential for holes $\mu_{\rm h}=11.4$ meV.~\cite{Kawamura,variation}
The calculated locations and heights of the peaks are in reasonable agreement with the experiment (except for the location of 1$\uparrow$, whose agreement is improved by the use of $B$-dependent chemical potential, see below). The good agreement infers that the phonon-drag is the dominant mechanism for the observed Nernst effect. 

Next we further include the contribution of electrons.
\noindent
Instead of using a fixed value, we now use a $B$-dependent chemical potential satisfying the charge neutral condition;
that is, we determine it such that the number of electrons and holes are equal. (Note that $\mu_\mathrm{e}+\mu_\mathrm{h}=\mu_0$, where $\mu_\mathrm{e}$ and $\mu_\mathrm{h}$ represent chemical potentials for electrons and holes, respectively, and  $\mu_0$ the band overlap.)
We evaluated the chemical potentials by Smith {\it et al.}'s method.~\cite{smith}
Smith {\it et al.} used a model proposed by Lax {\it et al.},~\cite{Lax,Lax2} where the conduction band becomes non-parabolic under the influence of a filled band just below it.
The energy of electron in the Lax model is represented by 
\begin{eqnarray}
E_{\rm e} \left( 1+ \frac{E_{\rm e}}{E_g}\right) = \hbar \omega_{\rm e} \left(n+\frac12 \right)+ \frac{\hbar^2 k_z^2}{2 m_{{\rm e}z}} + \sigma g_{\rm e} \mu_B B,
\label{Laxmodel}
\end{eqnarray}
where $E_g$ is the band gap between the conduction band and the filled band.
According to the model, the chemical potential for electrons (holes) increases (decreases) as the magnetic field is increased.
In the actual calculation of Eq.~(\ref{syxlast}), we linearized the Lax model (\ref{Laxmodel}) around $E_{\rm e}=\mu_{\rm e}$ for simplicity, noting that only the energy level in the immediate vicinity of the chemical potential is relevant at low temperatures. 



\begin{figure}
\begin{center}
\includegraphics[width=3.2in]{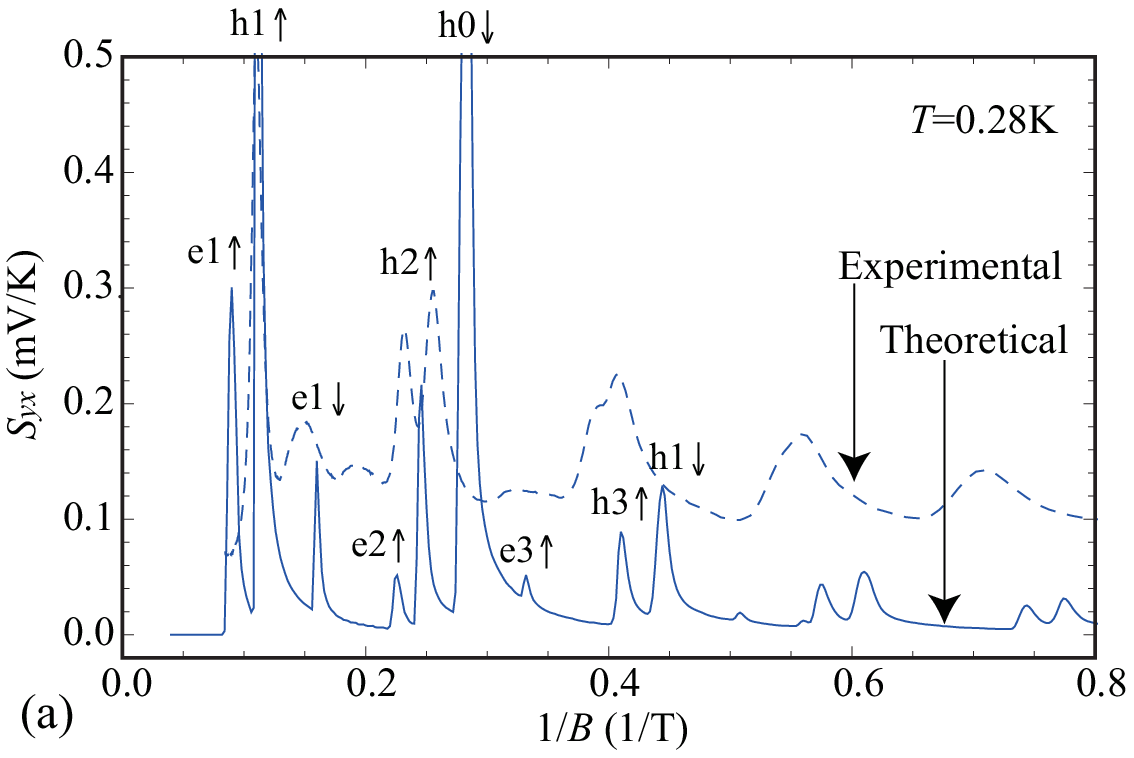}
\includegraphics[width=3.2in]{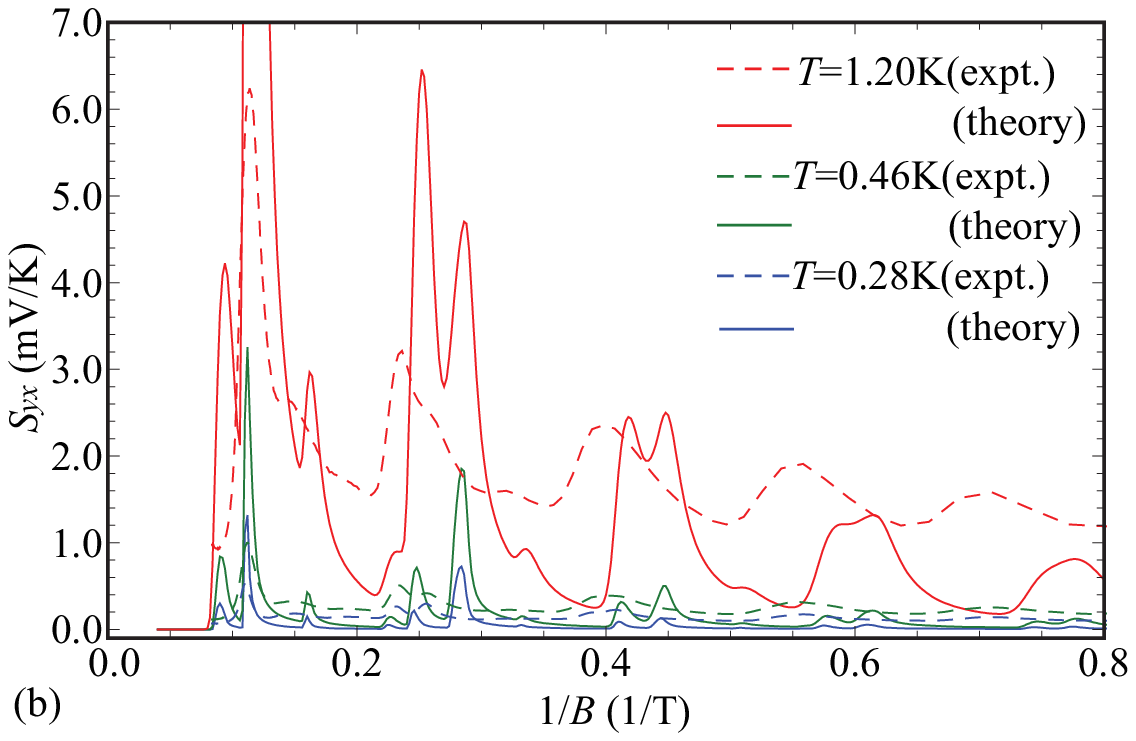}
\end{center}
\caption{(Color online). (a) The transverse thermopower $S_{yx}$ of holes and electrons at $T$= 0.28 K
 against the inverse magnetic field $1/B$.
(b) The data at $T$=0.28, 0.46 K, 1.20 K (from bottom to top). 
The solid lines are our theoretical results and the broken lines are the experimental results by Behnia {\it et al}.~\protect\cite{Behnia}. Note that the peak labeled  e2$\downarrow$ is overlapping with h0$\downarrow$. The peaks are labeled by $(a,n,{\rm spin})$.} 
\label{phonondragelec}
\end{figure}

The result is shown in Fig.\ \ref{phonondragelec}.
We again used the parameter values shown in Table \ref{param}. 
The peaks originating from holes remain basically unchanged from Fig.~\ref{phonondraghole}, but we now have
 additional peaks resulting from electrons. 
Our result suggests the possibility that the minor peaks between the major ones observed in the experiment are due to the electron contribution, rather than to the fractional quantization as implied in Ref.~\onlinecite{BehniaScience}. 
The peak labeled as e1$\uparrow$ may correspond to the peak at $14$ T shown in Fig.\ 1 
of Ref.~\onlinecite{BehniaScience} (not shown in Fig.\ \ref{phonondragelec}). 
The calculated values of $S_{yx}$ at the peaks due to electrons, as well as those of holes at low magnetic-field regime, are substantially smaller than those of the experiment. However, the height of the peaks, if we disregard the smooth background that considerably differs between the calculation and the experiment, are in rough agreement. The origin of the smooth background observed in the experiment is not known at present but is presumably related to the presence of disorders completely neglected in our calculation. Further discussion on the role of the disorder will be given below. 
In the calculation of electron contribution, we considered one of the three equivalent electron pockets rotated by $120^{\circ}$ to each other,~\cite{EdelmanReview}
one with the long axis parallel to the heat current, and simply multiplied the result by three, neglecting the anisotropy.
We estimate that the peak heights would become slightly smaller than those shown in Fig.~\ref{phonondragelec} 
due to the anisotropy, although it is difficult to take full account of the anisotropy in the calculation. 
Compared with Fig.~\ref{phonondraghole}, the peak h1$\uparrow$ shifted to lower magnetic-field side owing to decrease in $\mu_{\rm h}$ with increasing $B$, and coincide better with the experimental peak, while agreement of the positions of other major peaks slightly worsen. 
The slight inconsistency of the peak locations may be attributable to the minute discrepancy between values of the effective masses, the $g$ factor, and the band parameters in the literature and those of the sample used in the experiment.
(Very recently, it has been pointed out that slight misalignment in the direction of the magnetic field from the trigonal axis can also cause small shift in the peak positions.~\cite{Sharlai})
In Fig.~\ref{phonondragelec}(b) it can be seen that the strong temperature dependence of the experimental peak heights is reproduced well in the calculation. 



\section{Discussion}

We now comment on several characteristics of bismuth and/or the phonon-drag effect that are operative in yielding the sharp and large amplitude oscillation of $S_{yx}$.
(i) The small carrier density in bismuth is advantageous to the phonon-drag effect, since 
carriers with small Fermi momentum readily interact with phonons.
(ii) The conservation of energy and momentum in the carrier-phonon interaction, 
${\hbar^2}\left( k_z'^2 - k_z^2 \right)/{2 m_{az}}  = \hbar v_s q$ 
and
$k_z'=k_z + q_z$,  leads to $k_z = {\rm O} (q)$ [see Eq.\ (\ref{kz0})].
Here we consider only the intra-Landau-level scattering ($n = n'$); the inter-Landau-level scattering is practically prohibited in a quantizing magnetic field since $\hbar \omega_a \gg \hbar \omega_q$.
Since only phonons having small $q$ are available at low temperatures, 
only carriers with small $k_z$ are involved in the phonon-drag events, resulting in sharp peaks where Landau levels cross the chemical potential [see Eqs.\ (\ref{eq-energy}) and (\ref{syxlast})].
 (iii) The dominance of $S_{yx}$ over $S_{xx}$ is ascribable to the relation $\rho_{xx} \gg |\rho_{yx}|$ in bismuth, which contains both holes and electrons as carriers. 
The longitudinal and transverse thermopowers $S_{xx}$ and $S_{yx}$ are given by
$S_{xx} = \rho_{xx} \epsilon_{xx} - \rho_{yx} \epsilon_{yx}$ and $S_{yx} = \rho_{yx} \epsilon_{xx} + \rho_{xx} \epsilon_{yx}$, respectively, where $\boldsymbol{\epsilon}$ is the thermoelectric tensor.  
For the phonon-drag effect, it has been shown that   $|\epsilon_{yx}| \gg |\epsilon_{xx}| \sim 0$,~\cite{Fromhold} resulting in $|S_{yx}| \gg |S_{xx}|$ for bismuth, or for ambipolar conductor in general (and $|S_{yx}| \ll |S_{xx}|$ for 2DEGs or generally for systems with $\rho_{xx} \ll |\rho_{yx}|$); roughly speaking $S_{yx}$ in bismuth corresponds to $S_{xx}$ in 2DEGs.
The relation $|\epsilon_{xx}| \sim 0$ also allows us to evaluate $S_{xx}\simeq -\rho_{yx}\epsilon_{yx}$ simply by replacing $\rho_{xx}$ in Eq.\ (\ref{syxlast}) with $\rho_{yx}$. Using the experimentally obtained $\rho_{yx}$,~\cite{Behniadata} the calculation yields $S_{xx}$ having the peaks at roughly the same positions as in $S_{yx}$ but $\sim$1/20 in magnitude. The relation between $S_{yx}$ and $S_{xx}$ is in rough agreement with the experimental result shown in Fig.\ 1 of Ref.~\onlinecite{Behnia}.

We note in passing that a rather large fraction of the observed Nernst signal was ascribed to the diffusion contribution in Ref.\ \onlinecite{Behnia07-1} (see Fig.\ 2 in Ref.\ \onlinecite{Behnia07-1}), which appears to be in a mild contradiction to our conclusion.
We suspect, however, that the treatment described in their paper may not be estimating the magnitude of the diffusion contribution accurately for the following reasons:
(a) They used the relation between the Nernst coefficient and the Hall angle, Eq.\ (1) in their paper, which is not directly applicable to bismuth containing both electrons and holes with different Fermi energies.
(b) They seem to have used $\omega_c \tau (\gg1)$ as an estimate for the small Hall angle in bismuth (although they themselves seem to acknowledge the discrepancy between $\omega_c \tau$ and the Hall angle in bismuth).
(c) They replaced $\partial \tau / \partial \epsilon |_{\epsilon_{F}}$ by $\tau / \epsilon_{F}$, which, we think, is not readily justifiable. We consider, especially for (b), that the diffusion contribution can be smaller than their estimate.
Furthermore, their estimate is for a rather small magnetic $B=0.1$ T\@. In a quantizing magnetic field discussed in the present paper, the edge current (or surface diamagnetic current) should be taken into account, \cite{Obraztsov,nernst,Jonson84,Oji84} as we already mentioned in the introduction.


In our calculation, we neglected disorders in bismuth altogether. Although the effect of disorders is expected to be rather small in a high-quality bismuth single crystal, we believe that it constitutes the main source of the remnant discrepancy between the theoretical and the experimental traces. Inclusion of disorders introduces a width in the energy of Landau levels represented by the first term in Eq.\ (\ref{eq-energy}). The delta function in Eq.\ (\ref{Qydz}) denoting the energy conservation is then replaced by a peak function having the width acquired by the Landau levels, thereby making the peaks in $S_{yx}$ broader ~\cite{Kubakaddi,shirasaki} with concomitant decrease in the peak heights. The narrower peak width in the theoretical curves that allows some of the peaks not well-resolved in the experiment to be resolved is thus attributable to the neglect of the disorders in our calculation. The width in the Landau levels will also affect the kinetics involved in the carrier-phonon interaction. In the energy and momentum conservation mentioned above, only the kinetic energy in the $z$ direction, $\hbar^2 k_z^2/2 m_{a z}$, was allowed to vary, since the kinetic energy in the $x$-$y$ plane was strictly fixed to the Landau levels. Introduction of the width into the Landau levels alters the situation; the phonons can now also impart their energy to the in-plane kinetic energy of the carriers without affecting $\hbar^2 k_z^2/2 m_{a z}$. The restriction on the extent of $k_z$ mentioned above is thus removed, enabling the carrier-phonon scattering to take place regardless of the value of $k_z$. This may partly be responsible for the smooth background observed in the experiment.

\section{Conclusions}
We have calculated the transverse thermopower $S_{yx}$ due to the phonon-drag effect, taking both holes and electrons into account as carriers. A series of large ($\sim$ mV/K) peaks originating from holes, with smaller peaks deriving from electrons in between, are obtained. The heights as well as the positions of the peaks are close to those recently observed experimentally by Behnia \textit{et al},~\cite{Behnia} in stark contrast with the calculation based on the edge-current picture, corresponding to the diffusion contribution, in which the peak heights are orders of magnitude smaller. This strongly suggests that the phonon-drag is the dominant mechanism in the experimentally observed prominent magneto-oscillations in the Nernst coefficient. Rather broad width of the peaks and the smooth background not reproduced in our calculation are attributable to the disorders neglected in our calculation.

\begin{acknowledgements}
We thank Y.~Hasegawa for constructive comments, and
K.~Behnia for providing the experimental data.
The work is supported partly by the Thermal \& Electric Energy Technology Foundation, Foundation for Promotion of Material Science and Technology of Japan, and the Iketani Foundation
as well as by
NINS' Creating Innovative Research Fields Project (No.~NIFS08KEIN0091) and Grants-in-Aid for Scientific Research (Nos.~17340115 and 20340101) from The Ministry of Education, Culture, Sports, Science and Technology (MEXT). 
The computation was partly done using the facilities of Supercomputer Center, Institute for Solid State Physics, University of Tokyo, Japan.
\end{acknowledgements}


\begin{thebibliography}{99}
\bibitem{elemass}
Here we interchanged the $x$ and $y$ axes from the standard notation for the convenience in the later calculation.
\bibitem{variation} Slight variation between literatures is found on the value of $\mu_\mathrm{h}$. For example, $\mu_\mathrm{h} = 10.9$ meV is given in Ref.~\onlinecite{smith}.
\end{thebibliography}

\end{document}